\documentclass[12pt]{article}
\def\gappeq{\mathrel{\rlap {\raise.5ex\hbox{$>$}}
{\lower.5ex\hbox{$\sim$}}}}
\def\permil{$\%\raise.20ex\hbox{$_0$}}
\def\lappeq{\mathrel{\rlap{\raise.5ex\hbox{$<$}}
{\lower.5ex\hbox{$\sim$}}}}

\begin{document}
\topmargin -1.0cm
\oddsidemargin -0.8cm
\evensidemargin -0.8cm
\pagestyle{empty}
\begin{flushright}
UNIL-IPT-01-08\\
May 2001
\end{flushright}
\vspace*{5mm}

\begin{center}

{\Large\bf Localization of metric fluctuations on scalar branes}\\
\vspace{1.0cm}

{\large Massimo Giovannini\footnote{Electronic address: 
massimo.giovannini@ipt.unil.ch}}\\
\vspace{.6cm}
{\it {Institute of Theoretical Physics\\ University of Lausanne\\ 
CH-1015 Lausanne, Switzerland}}
\vspace{.4cm}
\end{center}

\vspace{1cm}
\begin{abstract}
The localization of metric fluctuations on scalar brane configurations 
breaking spontaneously five-dimensional Poincar\'e invariance is discussed. 
Assuming that the four-dimensional Planck mass is finite and that the
geometry is regular, it is demonstrated that  the vector and scalar 
fluctuations of the metric are not localized on the brane. 
\end{abstract}

\vfill

\eject
\pagestyle{empty}
\setcounter{page}{1}
\setcounter{footnote}{0}
\pagestyle{plain}


If internal dimensions are not compact \cite{RS1,RS2,akama,visser}  
all the fields describing 
the fundamental forces of our four-dimensional world should 
be localized on a higher dimensional topological 
defect \cite{RS1,RS2}. Localized means that 
the various fields should exhibit normalizable zero modes with respect 
to the bulk coordinates parameterizing the geometry of the 
defect in the extra-dimensional space. 
Among the various interactions an important  r\^ole is played by 
gravitational forces \cite{rusu1,rusu2}
 since the localization of the metric fluctuations 
can lead to computable and measurable deviations of Newton law at short 
distances \cite{exp}. In \cite{rusu2} it has been shown that the zero mode 
related to the tensor fluctuations of the geometry is localized, provided
the four-dimensional Planck mass is finite. For a recent 
review see \cite{rub}.

If the four-dimensional world is Poincar\'e-invariant, the higher-dimensional 
geometry will have not only tensor modes but also scalar and  vector 
fluctuations. The localization of the 
various modes of the geometry will then be the subject of the present 
analysis. On top of four-dimensional Poincar\'e invariance, 
the invariance of the fluctuations under infinitesimal coordinate 
transformations, i.e. gauge-invariance \cite{bardeen}, can be used 
in order to simplify the problem.
Gauge invariance guarantees that the equations for the fluctuations 
of the geometry do not change when moving from one coordinate system 
to the other. The tensor modes of the geometry are gauge-invariant 
but the scalar and vector modes are not. Hence, gauge-invariant
fluctuations corresponding to these modes should be constructed and analyzed.
The general derivation of these equations allows a model-independent
 discussion of the 
normalizability properties of the zero modes related to the various
fluctuations of the metric.

The following five-dimensional action \footnote{ Latin (uppercase) indices 
run over the five-dimensional space-time whereas Greek indices run over the 
four-dimensional space-time. Notice that 
 $\kappa = 8\pi G_{5} = 8\pi/M_{5}^3$. Natural gravitational 
units will be often employed by setting $2 \kappa =1$.}. 
\begin{equation}
S= \int d^{5}x \sqrt{|G|}\biggl[- \frac{R}{2\kappa} + \frac{1}{2} G^{A B} 
\partial_{A} \varphi \partial_{B} \varphi - V(\varphi)\biggr],
\label{ac}
\end{equation}
can be used in order to describe the breaking of five-dimensional Poincar\'e 
symmetry. Consider a potential which is invariant under the 
$\varphi \rightarrow -\varphi$ symmetry. Then, non-singular domain-wall 
solutions can be obtained, for various potentials, in a metric 
\begin{equation}
ds^2 = {\overline G}_{A B} dx^A d x^B  = a^2(w) [dt^2 - d\vec{x}^2 - dw^2].
\label{m1}
\end{equation}
For instance solutions of the type 
\begin{eqnarray}
&& a(w) = \frac{1}{\sqrt{b^2 w^2 + 1}},
\label{s1}\\
&& \varphi = \varphi(w) = \sqrt{6} \arctan{b w},
\label{s2}
\end{eqnarray}
is found for trigonometric potentials \cite{gr,kt,fr} like $V(\varphi) 
\propto ( 1 - 5 \sin^2{\varphi/\sqrt{6}})$. Solutions like Eqs. 
(\ref{s1})--(\ref{s2}) represent a smooth  version of the Randall-Sundrum
scenario \cite{rusu1,rusu2}.
The assumptions of the present analysis will now be listed. 

(i) The five-dimensional geometry is regular (in a technical sense)
for any value of the bulk coordinate $w$. This implies that singularities 
in the curvature invariants are absent.

(ii) Five-dimensional Poincar\'e invariance is broken 
through a smooth five dimensional domain-wall solution generated 
by a potential $V(\varphi)$ which is invariant under 
$\varphi \rightarrow - \varphi$. The warp factor $a(w)$ will 
then be assumed symmetric for $w\rightarrow - w$.

(iii) Four-dimensional Planck mass is finite because the following integral
converges
\begin{equation}
M^2_{P} \simeq M^3 \int_{-\infty}^{\infty} dw a^3(w).
\label{pl}
\end{equation}

(iv) Five-dimensional gravity is described according to Eq. (\ref{ac}) 
and, consequently, the equations of motion for the warped background 
generated by the smooth wall are, in natural gravitational units, 
\begin{eqnarray}
&&{\varphi'}^2 = 6 ( {\cal H}^2 - {\cal H}'), 
\label{c1}\\
&& Va^2 = - 3 ( 3 {\cal H}^2 + {\cal H}').
\label{c2}
\end{eqnarray}
where the prime denotes derivation with respect to $w$ and ${\cal H} = a'/a$.

Under these assumptions it will be shown that the gauge-invariant 
fluctuations corresponding to scalar and vector modes of the geometry 
are not localized on the wall. On the contrary, tensor modes of the geometry 
will be shown to be localized. This program will be achieved in two steps. 
In the first step decoupled equations for the gauge-invariant 
variables describing the fluctuations of the geometry will be obtained 
for general brane backgrounds
without assuming any specific solution.
In the second step the normalizability of the zero modes will be  
addressed using only the assumptions (i)--(iv).

Let us start by perturbing  the five-dimensional Einstein equations and 
of the equation for the scalar field  
\begin{eqnarray}
&&\delta R_{A B} = \frac{1}{2} \,\partial_{A} \varphi \partial_{B} \chi\, + 
\frac{1}{2} \partial_{A} \chi \partial_{B} \varphi\, - \frac{1}{3} 
\frac{\partial V}{\partial\varphi} \,\chi\, \overline{G}_{AB} - \frac{V}{3}\, 
\delta G_{A B},
\label{P1}\\
&& \delta G^{A B} \,\biggl( \partial_{A} \partial_{B} \varphi -
\overline{\Gamma}_{A B}^{C}\, \partial_{C} \varphi \biggr) 
+ \overline{G}^{A B} \,\biggl( \partial_{A} \partial_{B} \chi - 
\overline{\Gamma}_{A B}^{C} \partial_{C} \chi  - \delta 
\Gamma^{C}_{A B} \partial\varphi\biggr) 
+ \frac{\partial^2 V}{\partial \varphi^2}\chi=0,
\label{P2}
\end{eqnarray}
where the metric and the scalar field have been separated  
into their background and perturbation parts:
\begin{eqnarray}
&& G_{AB}(x^{\mu}, w) = \overline{G}_{AB}(w) + \delta G_{AB}(x^{\mu},w),
\nonumber\\
&& \varphi(x^{\mu}, w) = \varphi(w) + \chi(x^{\mu},w).
\label{chi}
\end{eqnarray}
In Eqs. (\ref{P1})-(\ref{P2}), $\delta \Gamma_{AB}^{C}$ and 
$\delta R_{A B}$ are, respectively, the fluctuations of the
Christoffel connections and of the  Ricci tensors, whereas 
$\overline{\Gamma}_{AB}^{C}$ are the values of the connections
computed using the background metric (\ref{m1}).
According to Eq. (\ref{P1}) and (\ref{P2}), the fluctuations of the 
scalar brane, $\chi$ are coupled to the scalar fluctuations of the geometry 
$\delta G_{AB}(x^{\mu}, w)$  whose modes can be decomposed 
using Poincar\'e invariance in four dimensions as
\begin{equation}
\delta G_{A B}= a^2(w) \biggl(\matrix{2 h_{\mu\nu} 
+(\partial_{\mu} f_{\nu} +\partial_{\nu} f_{\mu}) 
+ 2\eta_{\mu \nu} \psi
+ 2 \partial_{\mu}\partial_{\nu} E
& D_{\mu} + \partial_{\mu} C &\cr
D_{\mu} + \partial_{\mu} C  & 2 \xi &\cr} \biggr).
\label{pert}
\end{equation}
On top of $h_{\mu\nu}$ which is divergence-less and trace-less (i.e. 
$\partial_{\mu}h^{\mu}_{\nu}$=0, $h_{\mu}^{\mu}=0$) there are four scalars
(i.e. $E$, $\psi$, $\xi$ and $C$) and two divergence-less vectors 
($D_{\mu}$ and $f_{\mu}$). For infinitesimal coordinate transformations 
$x^{A} \rightarrow \tilde{x}^{A} = x^{A} + \epsilon^{A}$ the 
tensors $h_{\mu\nu}$ are invariant whereas the vectors and the scalars 
transform non-trivially. The four-dimensional part of the infinitesimal shift
$\epsilon_{A} = a^2(w)(\epsilon_{\mu}, - \epsilon_{w})$, can be decomposed as
 $\epsilon_{\mu}= \partial_{\mu} \epsilon + \zeta_{\mu}$ where 
$\zeta_{\mu}$ is a divergence-less vector and $\epsilon$  is a scalar. 

It is then clear that the 
 transformations for the scalars involve two gauge functions $\epsilon$ 
and $\epsilon_{w}$. The transformations for the vectors involve $\zeta_{\mu}$.
Two scalars and one divergence-less vector can be gauged-away 
by fixing the scalar and the vector gauge functions. In a different 
perspective \cite{bardeen}, since there are two scalar gauge functions and 
four scalar fluctuations of the metric (\ref{pert}), two gauge-invariant
(scalar) variables can be defined. In the present case the gauge-invariant 
scalar variables can be chosen to be:
\begin{eqnarray}
&&\Psi = \psi - {\cal H}  ( E' - C), 
\nonumber\\
&& \Xi =\xi - \frac{1}{a} [ a( C - E')]'.
\label{giscal}
\end{eqnarray}
By shifting infinitesimally the coordinate system from $x^{A}$ to 
$\tilde{x}^{A}$ the metric fluctuations change as 
\begin{equation}
\delta {G}_{AB}(x^{\mu},w) \rightarrow  \delta \tilde{G}_{A B} = 
\delta G_{AB} - \nabla_{A} \epsilon_{B} - \nabla_{B}
\epsilon_{A}, 
\label{liederiv}
\end{equation}
where the covariant derivatives are computed using the backround metric 
of Eq. (\ref{m1}). In spite of this,   
$\tilde{\Psi}= \Psi$, $\tilde{\Xi} = \Xi$ and $\tilde{h}_{\mu\nu} 
= h_{\mu\nu}$. The 
physical interpretation of $\Xi$ and $\Psi$ is clear if a specific gauge 
choice is selected. In the longitudinal gauge (i.e. $E=0$, $C=0$ and 
$f_{\mu}=0$) $\Xi= \xi$ and $\Psi =\psi$. Thus, in the longitudinal 
gauge, $\Psi$ and $\Xi$ are identical to the metric perturbations 
in a conformally Newtonian coordinate system. The 
scalar field fluctuation of Eq. (\ref{chi}) is not gauge-invariant and 
the gauge-invariant variable associated with it is:
\begin{equation}
X = \chi - \varphi' (E' -C).
\label{chigi}
\end{equation}

Since there is one vector gauge function,i. e. $\zeta_{\mu}$, 
one gauge-invariant variable 
can be constructed out of $D_{\mu}$ and $f_{\mu}$:
\begin{equation}
V_{\mu} = D_{\mu} - f_{\mu}'.
\label{givec}
\end{equation}
In terms of the variables defined in Eqs. (\ref{giscal})--(\ref{chigi}) 
and (\ref{givec}) the perturbed system of Eqs. (\ref{P1})--(\ref{P2}) 
can be written in a fully gauge-invariant way. The equation for the 
tensors, as expected, decouples from the very beginning:
\begin{equation}
\mu_{\mu\nu}'' - \partial_{\alpha}\partial^{\alpha} \mu_{\mu\nu} 
- \frac{(a^{3/2})''}{a^{3/2}} \mu_{\mu\nu} =0.
\label{mu}
\end{equation}
where $\mu_{\mu\nu} =a^{3/2} h_{\mu\nu}$ is the canonical 
normal mode of the of the action (\ref{ac}) perturbed to second order 
in the amplitude of tensor fluctuations.

The scalar variables (\ref{giscal}) and (\ref{chigi}) 
form a closed system consisting of the diagonal components of Eq. (\ref{P1}) 
\begin{eqnarray}
&&\Psi'' + 7 {\cal H} \,\Psi' + {\cal H}\, \Xi' + 
2 ( {\cal H}' + 3 {\cal H}^2)\,\Xi + \frac{1}{3} 
\frac{\partial V}{\partial\varphi} a^2 X - \partial_{\alpha}
\partial^{\alpha} \Psi =0,
\label{ps}\\
&&- \partial_{\alpha}\partial^{\alpha}\,\Xi - 4\, [ \Psi''+ {\cal H}\,\Psi'] - 
4 {\cal H}\, \Xi' - \varphi' X' - \frac{1}{3} 
\frac{\partial V}{\partial\varphi} a^2 X + \frac{2}{3} V a^2 \Xi =0,
\label{xi}\\
\end{eqnarray}
supplemented by the gauge-invariant version of the 
 perturbed scalar field equation (\ref{P2})
\begin{equation}
\partial_{\alpha} \partial^{\alpha} X - X'' - 3 {\cal H} X' + 
\frac{ \partial^2 V}{\partial\varphi^2} a^2 X - 
\varphi' [ 4 \Psi' + \Xi'] - 2 \Xi\,( \varphi'' + 3 {\cal H} \varphi') =0,
\label{ch}
\end{equation}
and subjected to the constraints 
\begin{eqnarray}
&&\partial_{\mu}\partial_{\nu}[ \Xi - 2 \Psi] =0,
\label{con1}\\
&& 6 {\cal H}\, \Xi + 6 \Psi' + X \,\varphi' =0.
\label{con3}
\end{eqnarray}
coming from the off-diagonal components of Eq. (\ref{P1}).
From Eq. (\ref{P1}), the 
 evolution of the gauge-invariant vector variable (\ref{givec}) is 
\begin{equation}
\partial_{\alpha} \partial^{\alpha} {\cal V}_{\mu} =0,\,\,\,\,\,\,
{\cal V}_{\mu}' + 
\frac{3}{2} {\cal H} {\cal V}_{\mu}=0,
\label{calV}
\end{equation}
where ${\cal V}_{\mu} = a^{3/2} V_{\mu}$ is the canonical 
normal mode of the action (\ref{ac}) perturbed to second order in the 
amplitude of vector fluctuations of the metric. 

Using repeatedly the constraints of Eqs. (\ref{con1})--(\ref{con3}), 
together with the background relations (\ref{c1})--(\ref{c2}), 
the scalar system can be reduced to the following two equations
\begin{eqnarray}
&& \Phi'' - \partial_{\alpha}\partial^{\alpha} \Phi - 
z\biggl(\frac{1}{z}\biggr)'' \Phi =0,
\label{ph}\\
&& {\cal G}'' - \partial_{\alpha}\partial^{\alpha}{\cal G}
- \frac{z''}{z} {\cal G} =0,
\label{G}
\end{eqnarray}
where 
\begin{equation}
\Phi= \frac{a^{3/2}}{\varphi'} \Psi,\,\,\,\,\,\,\,{\cal G} = 
a^{3/2} \Psi - z X.
\label{def}
\end{equation}
The same equation satisfied by $\Psi$ is also satisfied by $\Xi$ by virtue 
of the constraint (\ref{con1}).
In Eq. (\ref{G}) and (\ref{def}) the background dependence appears in terms 
of the ``universal'' function $z(w)$ 
\begin{equation}
z(w) = \frac{a^{3/2} \varphi'}{{\cal H}}.
\label{z}
\end{equation}
It should be appreciated that these equations are completely general
and do not depend on the specific background but only upon the general
form of the metric (\ref{m1}) and of the action (\ref{ac}). In fact, in order
to derive Eqs. (\ref{ph})--(\ref{G}) and (\ref{def})--(\ref{z}) 
no specific background has been assumed, but only Eqs. (\ref{c1})--(\ref{c2}) 
which come directly from Eq. (\ref{ac}) and hold for any choice of the 
potential generating the scalar brane configuration.

The effective ``potentials'' appearing 
in the Schr\"odinger-like equations (\ref{ph}) and (\ref{G}) are dual 
with respect to $z\rightarrow 1/z$. It can be shown that the 
 gauge-invariant function ${\cal G}$  is the normal mode of the  
action perturbed to second order 
in the amplitude of the scalar fluctuations of the metric analogous 
to the normal modes one can obtain in the case of compact internal 
dimensions \cite{mg} of Kaluza-Klein type.

Let us now discuss the localization of the zero modes of the various 
fluctuations and enter the second step of the present discussion.
The lowest mass eigenstate of Eq. (\ref{mu}) 
is  $\mu(w) = {\mu_0} a^{3/2}(w)$. Hence, the normalization 
condition of the tensor zero mode implies 
\begin{equation}
|\mu_0|^2 \int_{-\infty}^{\infty} a^3 ~dw = 2 |\mu_0|^2\int_{0}^{\infty}a^3(w)
~ dw=1 .
\end{equation} 
where the assumed $w\rightarrow -w $ symmetry of the background geometry 
has been exploited. Using assumptions (i), (ii) and (iii) the tensor 
zero mode is then normalizable \cite{rusu1,rusu2}. 

Let us now move to the analysis of vector fluctuations.
 Eq. (\ref{calV}) shows that the vector fluctuations are 
always massless and the corresponding zero mode is ${\cal V}(w) \sim 
{\cal V}_0 a^{-3/2} $. Consequently, the normalization 
condition will be 
\begin{equation}
2 |{\cal V}_0|^2 \int_{0}^{\infty} \frac{d w}{a^3(w)} =1,
\end{equation}
which cannot be satisfied if assumption (i), (ii) and (iii) hold. If $a^3(w)$ 
converges everywhere, $1/a^3(w)$ will not be convergent.
Therefore, if the four-dimensional Planck mass 
is finite the tensor modes of the geometry are normalizable and the vectors 
are not.

From Eq. (\ref{ph}) the lowest mass eigenstate 
of the metric fluctuation $\Phi$ corresponds to 
$\Phi(w)= \Phi_0 z^{-1}(w)$ and the 
related normalization condition reads 
\begin{equation} 
2|\Phi_0|^2 \int_{0}^{\infty} \frac{dw}{z^2(w)}=1.
\label{norm}
\end{equation}
The integrand appearing in Eq. (\ref{norm}) will now be shown 
to be non convergent at infinity if the geometry is regular. In fact 
 according to assumption (i)
\begin{eqnarray}
&& R = \frac{4}{a^2} ( 2 {\cal H}' + 3 {\cal H}^2),
~~~~~R_{M N}R^{M N} = \frac{4}{a^4}( 4 {\cal H}^4 + 6 {\cal H}' {\cal H}^2
+ 5 {{\cal H}'}^2),
\nonumber\\
&& R_{M N A B}R^{M N A B} = \frac{8}{a^4}( 
2 {{\cal H}'}^2 - 5 {\cal H}^4),
\label{curv}
\end{eqnarray}
should be regular for any $w$ and, in particular, at infinity. The absence 
of poles in the curvature invariants guarantees the regularity of the 
five-dimensional geometry. 
Eq. (\ref{curv}) rules then out, in the coordinate system of (\ref{m1}), 
warp factors decaying at infinity as $e^{-d w}$ or $e^{- d^2 w^2}$: these
profiles would lead to divergences in Eqs. (\ref{curv}) at infinity 
\footnote{In order to avoid confusions it should be stressed that 
exponential warp factors naturally appear in non-conformal coordinate systems
related to the one of Eq. (\ref{m1}) as $a(w) dw = dy$.}.
Since $a(w)$ must converge at infinity,  $a(w) \sim w^{-\gamma}$
with $1/3\leq\gamma \leq 1$. Notice that $\gamma\geq 1/3$ 
comes  from the
 convergence (at infinity) of the integral of Eq. (\ref{pl}) \footnote{ 
In this sense the power $\gamma$ measures only the degree of convergence 
of a given integral.} and that $\gamma \leq 1$ is implied 
by Eqs. (\ref{curv}) since, at infinity, 
$R_{M N}R^{M N} \sim R_{M N A B} R^{M N A B } \sim 
w^{4 ( \gamma - 1)}$ should converge.
Using  Eq. (\ref{z}) and Eq. (\ref{c1}) the integrand of Eq. (\ref{norm})
can be written as 
\begin{equation}
\frac{1}{z^2} = \frac{{\cal H}^2}{a^3 {\varphi'}^2} = \frac{1}{6a^3}
\biggl(\frac{{\cal H}^2}{ {\cal H}^2 - {\cal H}'}\biggr).
\label{1/z}
\end{equation}
The behavior at infinity of Eq. (\ref{1/z}) can be now investigated assuming
the regularity of Eqs. (\ref{curv}), i.e. 
 $a(w) \sim w^{-\gamma}$ with $0< \gamma \leq 1$. In this limit 
\begin{equation}
\lim_{w\rightarrow \infty} \biggl(\frac{{\cal H}^2}{{\cal H}^2 
- {\cal H}'} \biggr)\sim \frac{\gamma^2}{\gamma^2 -\gamma} .
\label{lim1}
\end{equation}
and $1/z^2$ diverges {\em at least} as $a^{-3}$. In fact, if  $\gamma =1$, 
$1/z^2$ diverges even more as it can be argued from Eq. (\ref{lim1}) which 
has a further pole for $\gamma^2 = \gamma$. 
The  example given in Eqs. (\ref{s1})--(\ref{s2}) corresponds to 
a behavior at infinity given by $\gamma=1$.  
Direct calculations show that $1/z^2$ diverges, in this case,  as $w^5$.

Consequently, if the four-dimensional Planck mass is finite and if 
space-time is  regular the gauge-invariant (scalar) 
 zero mode is not normalizable and not localized on the brane.
For sake of completeness it should be mentioned that, for the lowest 
mass eigenvalue, 
there is a second (linearly independent) solution to Eq. (\ref{ph}) 
which is given by $ z^{-1}(w) \int^{w} z^2(x) ~dx$ which has poles at 
infinity and for $w\rightarrow 0$.
The poles appearing for $w\rightarrow 0$
will now be discussed since they are needed in order to prove that 
the zero modes of Eq. (\ref{G}) are not localized. As far as the poles 
 at infinity are concerned it is interesting to consider 
what happens to  
$ z^{-1}(w) \int^{w} z^2(x) ~dx$ in the case of the solution 
(\ref{s1})--(\ref{s2}). In this case, by direct use of Eqs. 
(\ref{s1})--(\ref{s2}) and (\ref{z}) we have that the second solution
diverges, at infinity, as $ (1 + b^2 w^2)^{1/4}\, (1 + 2 b^2 w^2)$. 

Noticing the duality connecting the effective 
potentials of Eqs. (\ref{ph}) and (\ref{G}) it can be
verified that the lowest mass eigenstate of Eq. (\ref{G}) is 
given by ${\cal G}(w)= {\cal G}_0 z(w)$.  
Provided the assumptions (i)--(iv) are satisfied,
it will now be demonstrated that the integral
\begin{equation}
2 |{\cal G}_{0}|^2 \int_{0}^{\infty} z^2 ~dw,
\end{equation}
is divergent not because of the behavior  at infinity
but because of the behavior of the solution close to the core 
of the wall, i.e. $w\rightarrow 0$. 
Bearing in mind Eq. (\ref{curv}),
assumption (i) and (ii) imply that $a(w)$ and $\varphi$ 
should be regular for any $w$. More specifically
close to the core of the wall $\varphi$ should go to zero 
and  $a(w)$ should go to a 
constant because of $w\rightarrow - w$ 
symmetry and the following regular expansions can be written
for small $w$
\begin{eqnarray}
&& a(w) \simeq a_0 - a_1\, w^{\beta} + ..., ~~~~~\beta>0,
\label{exp1}\\
&& \varphi(w) \simeq \varphi_1\,w^{\alpha} +..., ~~~~~\alpha >0,
\label{exp2}
\end{eqnarray}
for $w\rightarrow 0$. Inserting 
the expansion (\ref{exp1})--(\ref{exp2}) 
into Eq. (\ref{c1}) the relations can be obtained:
\begin{equation}
\beta = 2\alpha, \,\,\,\, \alpha^2 \varphi_1^2 = 6 \frac{a_1}{a_0} 
~\beta(\beta-1).
\label{cond}
\end{equation}
Inserting now Eqs. (\ref{exp1})--(\ref{exp2}) into Eq. (\ref{z}) 
and exploiting the first of Eqs. (\ref{cond}) we have
\begin{equation}
\lim_{w\rightarrow 0} z^2(w) \simeq w^{2 (\alpha - \beta)} = w^{-2\alpha}.
\,\,\,\alpha >0
\label{divv}
\end{equation}
Using Eqs. (\ref{exp1}) into 
Eqs. (\ref{curv}), $R_{A B} R^{A B} \sim 
R_{M N A B} R^{ M N A B} \sim w^{ 2 \beta -4}$, which implies $\beta \geq 2$ 
in order to have regular invariants for $w \rightarrow 0$.  
Since, from Eq. (\ref{cond}),  $\beta = 2\alpha$, in Eq. (\ref{divv})  
it must be $\alpha \geq 1$.
As in the case of Eq. (\ref{ph}) also eq. (\ref{G}) has a second 
(linearly independent) solution for the lowest mass eigenvalue, namely 
$z(w) \int^{w} dx~z^{-2}(x)$ which has poles at infinity.  
In fact, a direct check shows that, at infinity, this quantity
 goes ar $w^{\frac{3}{2}\gamma+ 1}$ where, as usual, 
$1/3\leq\gamma\leq 1$ for the convergence of the Planck mass and of the 
curvature invariants at infinity. 

In conclusion, it has been demonstrated that under 
assumptions (i)--(iv), the scalar and vector fluctuations
 of the five-dimensional metric decouple from the wall. 
 Heeding experimental tests \cite{exp}, 
the present results suggest that, under the 
assumptions (i)--(iv), no vector or scalar component of the Newtonian 
potential at short distances should be expected.

It is pleasure to thank M. E. Shaposhnikov for important comments.

\end{document}